\documentclass[12pt]{article}

\usepackage{cite}
\usepackage{epsf}
\usepackage{graphicx}
\usepackage{psfrag}
\usepackage[dvips]{epsfig}

\usepackage[english]{babel}

\usepackage{amssymb,indentfirst}
\usepackage{amsmath}

\usepackage{verbatim}

\makeatletter
\renewcommand \thesection {\@arabic\c@section.}
\renewcommand\thesubsection   {\thesection\@arabic\c@subsection.}
\renewcommand\thesubsubsection{\thesubsection\@arabic\c@subsubsection.}
\makeatother

\def\starup#1{\mbox{$\raise1.8ex\hbox{$*$} \kern-.7em#1$}}
\def\krup#1{\mbox{$\raise1.8ex\hbox{$+$} \kern-1.0em#1$}}
\def\linup#1{\mbox{$\raise1.9ex\hbox{---} \kern-1.0em#1$}}

\begin{document}
\title{ The chiral color symmetry of quarks
\\ and $G'$-boson contributions to charge asymmetry
\\ in $t \bar t$-production at the LHC and Tevatron
}
\author{I.~V.~Frolov\footnote{{\bf e-mail}: pytnik89@mail.ru},
M.~V.~Martynov\footnote{{\bf e-mail}: martmix@mail.ru},
A.~D.~Smirnov\footnote{{\bf e-mail}: asmirnov@uniyar.ac.ru}
\\
{\small Division of Theoretical Physics, Department of Physics,}\\
{\small Yaroslavl State University, Sovietskaya 14,}\\
{\small 150000 Yaroslavl, Russia.}}
%\address{}%
%\email{}%
\date{}
\maketitle
%\thanks{}%

%\vspase
% ----------------------------------------------------------------

\begin{abstract}
\noindent
The contributions of $G'$-boson predicted by the chiral color symmetry of quarks
to the charge asymmetry $A_C(p p \to t\bar{t})$ in $t \bar{t}$ production at the LHC
and to the forward-backward asymmetry $A_{\rm FB}(p\bar{p} \to t\bar{t})$
in $t\bar{t}$ production at the Tevatron
are calculated and analysed in dependence on two free parameters of the model,
the $G'$ mass $m_{G'}$ and mixing angle $\theta_G$.
The $m_{G'} - \theta_G$ regions of $1 \sigma$ consistency with the CMS data on
the cross section $\sigma(pp \to t\bar{t})$ and on the charge asymmetry $A_C(p p \to t\bar{t})$
are found and compared with those resulted from the CDF data
on the cross section~$\sigma(p\bar{p} \to t\bar{t})$
and on the forward-backward asymmetry~$A_{\rm FB}(p\bar{p} \to t\bar{t})$
of $t \bar{t}$ production at the Tevatron
with account of the current SM predictions for~$A_{\rm FB}(p\bar{p} \to t\bar{t})$.

\vspace{5mm}


\end{abstract}

%\newpage

%-----------------------------------------------------------------

%\setlength{\baselineskip}{24pt}

The Standard Model (SM) of electroweak and strong interactions based on the gauge symmetry group
\begin{eqnarray}
G_{SM}=SU_c(3) \times SU_L(2) \times U(1)
\label{eq:GSM}
%\nonumber
\end{eqnarray}
\noindent
describes well the interactions of guarks and leptons and gauge fields at the energies
of order of hundreds GeV and is still consistent with all the experimental data,
including the current experimental data from LHC.
After discovery of the Higgs-like boson $H$ with mass $m_H \approx 126~GeV$
% \cite{:2012gk}
the investigations of the propeties of this boson as well as the further search
for the possible effects of new physics are the main goals of the experiments at the LHC.
There many models predicting new physics effects at the LHC (such as two Higgs models,
models with the fourth fermion generation, models based on supersymmetry, left-right symmetry,
four color quark-lepton symmetry, etc.) and the unobservation of these effects at the LHC will
set the new limits on the parameter of these models.

One of such models which also can be verified at the LHC is based on the idea of the originally
chiral character of $SU_c(3)$ color symmstry of quarks i.e on the gauge group of the chiral color symmetry
\begin{eqnarray}
\label{chiral_group}
%G_c=SU_L(3)\!\times \! SU_R(3) \to SU_c(3),
G_c = SU_L(3)\!\times \! SU_R(3)  \stackrel{\,M_{chc}\,}{\longrightarrow}  SU_c(3)
\end{eqnarray}
which is assumed to be valid at high energies and is broken to usual QCD $SU_c(3)$ at some low energy
scale $M_{chc}$.
It would be of interest to know the lower limit on this energy scale of the chiral color symmetry breaking
which is achieved now at the LHC.

The immediate consequence of the chiral color symmetry of quarks is the prediction of
a new gauge particle -- the axigluon \; $G^A$
(in particular case of $g_L=g_R$ \cite{Pati:1975ze,Hall:1985wz,Frampton:1987ut,Frampton:1987dn})
or $G'$-boson (in more general case of  $g_L\neq g_R$
\cite{Cuypers:1990hb,Martynov:2009en,Martynov:2010ed,Martynov:2012zz}).
The $G'$-boson is the color octet massive gauge particle with vector and axial
vector couplings to quarks of order $g_{st}$ defined by the gauge group~\eqref{chiral_group}.
As a result, the $G'$-boson can give rise to the increase
of the cross section and to the charge asymmetry of $t \bar t$ production
at the LHC as well as to the forward-backward asymmetry
in $t \bar t$ production at the Tevatron.
The possible effect of $G'$-boson on the forward-backward asymmetry
in $t \bar t$ production at the Tevatron has been considered
in refs.\cite{Martynov:2009en,Martynov:2010ed,Martynov:2012zz}.
The other possible explanations of the large forward-backward asymmetry
in $t \bar t$ production at the Tevatron are also known and discussed
in a large number of papers, see for instance 
recent papers~\cite{Kamenik:2011wt,Berger:2011sv,
Cao:2011hr,Chivukula:2011ng,Biswal:2012mr,Cvetic:2012kv,Zhu:2012um,Gabrielli:2011zw,Wang:2012zv}
and the references therein.

The charge asymmetry in $t \bar t$ production at the LHC has been measured
by CMS~\cite{Chatrchyan:2011hk, CMS-PAS-TOP-11-030, CMS-PAS-TOP-12-010 }
and ATLAS~\cite{ATLAS-CONF-2011-106, ATLAS-CONF-2012-057} Collaborations in agreement
(within experimental errows) with SM predictions.
For the further analysys we use
%
%The current
 the CMS data on cross section $\sigma_{t\bar{t}}$~\cite{CMS-PAS-TOP-11-024}
and charge asymmetry~$A_{\rm C}$~\cite{CMS-PAS-TOP-11-030}
 in the $t\bar{t}$ production at the LHC %are
\begin{eqnarray}
%\sigma_{t\bar{t}} & = & 7.5 \pm 0.31 (stat) \pm 0.34 (syst) \pm 0.15 (lumi) pb \, (= 7.5 \pm 0.48 \, pb) ,
\hspace{-10mm} && \sigma_{t\bar{t}} = 165.8 \pm 2.2(stat.)\pm 10.6(syst.)\pm 7.8(lumi.) \text{pb}
(=165.8 \pm 13.3 \text{pb}),
% \, \cite{Kidonakis2010}
%\\
\label{csppttbarCMS}
%\nonumber
\\
%A_{\rm FB}^{p \bar p} & = & 0.193 \pm 0.065~(\rm{stat}) \pm 0.024~(\rm{sys})\, (= 0.193 \pm 0.069),
%\;\;\;\; ( ?! )
%A_C &=& -0.013\pm 0.028(stat.) _{-0.031}^{+0.029}(syst.) \, \cite{CMSCollaboration2012}
%\nonumber
%\\
\hspace{-10mm} && A_C = 0.004\pm 0.010(stat.) \pm 0.012 (syst.)
(=0.004 \pm 0.015)
% \,  \cite{Coll.2012}
\label{AcppttbarCMS}
%\nonumber
\end{eqnarray}
and the SM predictions for $\sigma_{t\bar{t}}$~\cite{Kidonakis:2010dk}
and $A_{\rm C}$~\cite{Kuhn:2011ri}
\begin{eqnarray}
%\sigma_{t\bar{t}}^{SM} & = & 7.35 {~}^{+0.38}_{-0.80}~\mathrm{(scale)}
%{~}^{+0.49}_{-0.34} ~\mathrm{(PDFs)} \mathrm{[CTEQ6.5]}  \, \mathrm{pb}  \div
\sigma_{t\bar{t}}^{\text{NNLOapprox}}%(m_t &=&173 \text{GeV})
&=& 163_{-10}^{+11} \, \text{pb} \, ,
%\cite{Kidonakis2010}
% \nonumber
%  \\
\label{csppttbarSM}
%\nonumber
\\
%\label{sppttSM}
%\notag &\phantom{\div}& 7.93 {~}^{+0.34}_{-0.56}~\mathrm{(scale)}
%{~}^{+0.24}_{-0.20} ~\mathrm{(PDFs)} \mathrm{[MRST2006nnlo]} \, \mathrm{pb} ,
%\\
%\label{sppttSM}
%\\
%  A_{\rm FB}^{SM}(p \bar p \to t \bar t )  &=&  0.051(6), \;\;\;\; ( ?! )
A_C^{\text{SM}} &=& 0.0115\pm 0.0006.  % \, \cite{Kuhn2012}
\label{AcppttbarSM}
%\nonumber
\end{eqnarray}

In the present paper we calculate the contributions of the gauge $G'$-boson
to the cross section and to the charge asymmetry in the $t \bar{t}$ production at the LHC.
We compare the results with the CMS data~\eqref{csppttbarCMS},~\eqref{AcppttbarCMS}
and discuss the corresponding allowed region of the free parameters of the model
in comparision with that resulted from Tevatron data on the forward-backward asymmetry.

%\label{csppttbarCMS}

The interaction of the $G'$-boson with quarks defined by gauge chiral color symmetry
group~\eqref{chiral_group} can be written as
%in model independent way as
%
\begin{equation}
\mathcal{L}_{G'qq} =  \bar{q} \gamma^\mu (g_V + g_A \gamma_5) G'_\mu q =
g_{st}(M_{chc}) \, \bar{q} \gamma^\mu (v + a \gamma_5) G'_\mu q
\nonumber
\end{equation}
where $G'_{\mu} = G'^i_{\mu} t_i$, $t_i$, $i=1,2,...,8$  are the generators of $SU_c(3)$ group,
\begin{eqnarray}
g_{st}(M_{chc}) = \frac{g_L g_R}{\sqrt{(g_L)^2+(g_R)^2}}
\nonumber
\end{eqnarray}
is the strong interaction coupling constant
at the mass scale $M_{chc}$ of the chiral color symmetry breaking
and the vector and axial-vector coupling constants $v$ and $a$ are defined
by one parameter, the $G^{L} - G^{R}$ mixing angle $\theta_G$ as
\begin{eqnarray}
v = \frac{c_G^2-s_G^2}{2 s_G c_G} = \cot(2\theta_G), \,\,\,\,
a = \frac{1}{2 s_G c_G} = 1 / \sin(2\theta_G),
\label{eg:va}
\nonumber
\end{eqnarray}
 $s_G =\sin\theta_G$, $c_G =\cos\theta_G$,  $tg\,\theta_G=g_R/g_L$,
$g_L, \, g_R $ are the gauge coupling constants of the chiral color group~\eqref{chiral_group}.
As a result the gauge chiral color symmetry model has two free parameters, $G'$-boson mass $m_{G'}$ and
the $G^{L} - G^{R}$ mixing angle $\theta_G$. % , $tg\,\theta_G=g_R/g_L$.

In the parton process $q\bar{q} \rightarrow  t \bar{t}$ of $t \bar{t}$ production in $q \bar{q}$ collisions
%along the $z$ axis
the momenta of initial quarks in collinear limit can be written as
\begin{eqnarray}
  p_q &=& \{\varepsilon_q,0,0,p_{qz}\}, \hspace{10mm}
p_{\bar{q}} = \{\varepsilon_{\bar{q}},0,0,p_{\bar{q}z}\}
%\hspace{10mm}   \hat{s}=(p_q+p_{\bar{q}})^2
\nonumber
\label{eq:pqpq1}
\end{eqnarray}
with $\hat{s}=(p_q+p_{\bar{q}})^2$.

The momenta of the final $t$- and $\bar{t}$-quarks can be %written as
expressed in terms of
their rapidities $y_t, \,y_{\bar{t}}$ and of the transversal momentum $p_{\perp x}, \, p_{\perp y}$  as
\begin{eqnarray}
  p_t &=& \left\{\sqrt{p_{\perp}^2+m_t^2\,}\mathrm{ch}\,  y_t,\, p_{\perp x}, \, p_{\perp y}, \,
\sqrt{p_{\perp}^2+m_t^2\,}\mathrm{sh}\,  y_t\right\},
\nonumber
  \\
  p_{\bar{t}} &=& \left\{\sqrt{p_{\perp}^2+m_t^2\,}\mathrm{ch}\,  y_{\bar{t}},\,
- p_{\perp x}, \, - p_{\perp y}, \,\sqrt{p_{\perp}^2+m_t^2\,}\mathrm{sh}\,  y_{\bar{t}}\right\},
\nonumber
  \label{ptpt1}
\end{eqnarray}
where
\begin{eqnarray}
  y_i=\frac{1}{2}\ln\left(\frac{\varepsilon_i+p_{iz}}
{\varepsilon_i-p_{iz}}\right), \quad i=t, \bar{t}
  \nonumber
\label{rapi}
\end{eqnarray}
are the rapidities %$y_t, \,y_{\bar{t}}$
of $t$- and $\bar{t}$-quarks.
%and $p_{\perp x}, \, p_{\perp y}$ is the the transversal momentum of $t$-quark.
%

Below instead of the rapidities %~\eqref{rapi}
we  use the variables
%It is convinient below  instead of the rapidities %~\eqref{rapi}
%to use the variables
%
\begin{eqnarray}
Y &=& y_t+y_{\bar{t}}, \hspace{10mm}    z = y_t^2-y_{\bar{t}}^2
\nonumber
  \label{eq:2}
\end{eqnarray}
(one often uses the variable $\Delta |y| = |y_t| - |y_{\bar{t}}| $ instead of the variable $z$
but the variable $z$ is more convinient for mathematical manipulations).

The conservation of the 4-momentum
%
%$$p_q+p_{\bar{q}}=p_t+p_{\bar{t}}$$
%
fixes for $Y$  and $p_{\perp}$ the values
\begin{eqnarray} \overline{Y}=
\ln\left(\frac{\varepsilon_q+\varepsilon_{\bar{q}}+p_{qz}+
p_{\bar{q}z}}{\varepsilon_q+\varepsilon_{\bar{q}}-(p_{qz}+
p_{\bar{q}z})}\right),
\;\; \bar{p}_{\perp}^2+m_t^2 =
\frac{\hat{s}}{4\mathrm{ch}\, ^2(z/2\overline{Y})}
\nonumber
  \label{eq:8}
\end{eqnarray}
%%
%\begin{eqnarray}
%  \bar{p}_{\perp}^2+m_t^2 = \frac{\hat{s}}{4\mathrm{ch}\, ^2(z/2\overline{Y})},
% \nonumber
%  \label{eq:7}
%\end{eqnarray}
%%
whereas $z$ varies in interval
\begin{eqnarray}
 % &f(p_q,p_{\bar{q}})=p_{qz}\varepsilon_{\bar{q}}-p_{\bar{q}z}\varepsilon_q; % \nonumber
 % \\
  -z_0\le & z & \le z_0, \hspace{10mm} z_0 = \overline{Y}\Delta y_0,
%\nonumber
%  \\
%[4mm]
%  \mathrm{th}(\Delta y_0/2) &=& \sqrt{1-4m_t^2/\hat{s}}\equiv \beta.
\hspace{10mm} \mathrm{th}(\Delta y_0/2) = \sqrt{1-4m_t^2/\hat{s}}\equiv \beta.
\nonumber
  \label{eq:10}
\end{eqnarray}
%%
%\begin{eqnarray}
%  \hat{s}=(p_q+p_{\bar{q}})^2.
%\nonumber
%  \label{eq:4}
%\end{eqnarray}
%%

The total parton cross section of the process
$q\bar{q} \stackrel{\,g,\,G'}{\rightarrow} t \bar{t}$
%$q\bar{q} \rightarrow  t \bar{t}$
with account of the $G'$-boson
for $ m_q^2 \ll m_t^2,\, \hat{s}$ can be written as
\begin{eqnarray}
  \nonumber
%\sigma^{LO}(q\bar{q} \stackrel{\,g,\,G'}{\rightarrow} t \bar{t}) &=&
%\sigma^{LO}_{SM}(q\bar{q} \rightarrow t \bar{t}) + \Delta\sigma^{LO}_{G'}(q\bar{q} \rightarrow t \bar{t}),
\sigma(q\bar{q} \stackrel{\,g,\,G'}{\rightarrow} t \bar{t}) &=&
\sigma^{SM}(q\bar{q} \rightarrow t \bar{t}) +
\Delta\sigma_{LO}^{G'}(q\bar{q} \rightarrow t \bar{t})
%\\
%[4mm]
%% \nonumber
%%\sigma^{LO}_{SM}(q\bar{q} \rightarrow t \bar{t}) &=&
%%\frac{4\pi \beta}{27\hat{s}} \, \alpha_s^2(\mu) \, (3-\beta^2),
%%\\
%%[4mm]
% \nonumber
%%
%\Delta\sigma_{LO}^{G'}(q\bar{q} \rightarrow t \bar{t}) &=&
%\frac{4\pi \beta}{27}
%  \bigg \lbrace
%%\frac{4\pi \beta}{27\hat{s}}
%%  \bigg \lbrace
%%  \alpha_s^2(\mu) \, (3-\beta^2) +
%%\\
%% \nonumber
%%  &+&
%\frac{2\, \alpha_s(\mu)  \alpha_s(M_{chc})\, v^2 (\hat{s}-m_{G'}^2)(3-\beta^2)}
%  {(\hat{s}-m_{G'}^2)^2+\Gamma_{G'}^2 m_{G'}^2}+
%  \\
%  &+&\frac{\alpha_s^2(M_{chc}) \, \hat{s} (v^2+a^2) \big[ v^2(3-\beta^2) +2a^2\beta^2 \big]}
%  {(\hat{s}-m_{G'}^2)^2 + \Gamma_{G'}^2 m_{G'}^2}
%  \bigg \rbrace.
% \nonumber
%  \label{eq:17}
\end{eqnarray}
where $\sigma^{SM}(q\bar{q} \rightarrow t \bar{t})$ is the total parton cross section in the SM and
\begin{eqnarray}
%  \nonumber
%%  \sigma^{LO}(q\bar{q} \stackrel{\,g,\,G'}{\rightarrow} t \bar{t}) &=&
%%\sigma^{LO}_{SM}(q\bar{q} \rightarrow t \bar{t}) + \Delta\sigma^{LO}_{G'}(q\bar{q} \rightarrow t \bar{t}),
%  \sigma(q\bar{q} \stackrel{\,g,\,G'}{\rightarrow} t \bar{t}) &=&
%\sigma^{SM}(q\bar{q} \rightarrow t \bar{t}) +
%\Delta\sigma_{LO}^{G'}(q\bar{q} \rightarrow t \bar{t}),
%\\
%[4mm]
%% \nonumber
%%\sigma^{LO}_{SM}(q\bar{q} \rightarrow t \bar{t}) &=&
%%\frac{4\pi \beta}{27\hat{s}} \, \alpha_s^2(\mu) \, (3-\beta^2),
%%\\
%%[4mm]
% \nonumber
%%
\Delta\sigma_{LO}^{G'}(q\bar{q} \rightarrow t \bar{t}) &=&
\frac{4\pi \beta}{27}
  \bigg \lbrace
%\frac{4\pi \beta}{27\hat{s}}
%  \bigg \lbrace
%  \alpha_s^2(\mu) \, (3-\beta^2) +
%\\
 \nonumber
%  &+&
\frac{2\, \alpha_s(\mu)  \alpha_s(M_{chc})\, v^2 (\hat{s}-m_{G'}^2)(3-\beta^2)}
  {(\hat{s}-m_{G'}^2)^2+\Gamma_{G'}^2 m_{G'}^2}+
  \\
  &+&\frac{\alpha_s^2(M_{chc}) \, \hat{s} (v^2+a^2) \big[ v^2(3-\beta^2) +2a^2\beta^2 \big]}
  {(\hat{s}-m_{G'}^2)^2 + \Gamma_{G'}^2 m_{G'}^2}
  \bigg \rbrace
% \nonumber
\label{DsG1qqbarttbar}
\end{eqnarray}
is the contribution induced by the $G'$-boson in tree approximation,
$\mu$ is a typical energy scale of the process.

We define the charge difference of the parton cross sections
of the process $q\bar{q} \rightarrow  t \bar{t}$ as
\begin{eqnarray}
\Delta_C(q\bar{q}\to t\bar{t}) &=& \sigma( q\bar{q}\to t\bar{t}, z>0 )-\sigma( q\bar{q}\to t\bar{t}, z<0 ).
%\nonumber
\label{eq:DC}
\end{eqnarray}

We have found the $G'$-boson contribution of tree approximation to the charge difference
of the process
$ q\bar{q}\stackrel{\,g,\,G'}{\longrightarrow} t \bar{t} $
for $ m_q^2 \ll m_t^2,\, \hat{s} $
in the form
\begin{eqnarray}
  \Delta_C^{G'}(q\bar{q}\stackrel{\,g,\,G'}{\rightarrow} t\bar{t}) &=&
%  \Delta_C^{LO}(q\bar{q}\stackrel{\,g,\,G'}{\rightarrow} t\bar{t}) &=&
%\sigma(q\bar{q}\stackrel{\,g,\,G'}{\rightarrow}t\bar{t}, z>0) -
%\sigma(q\bar{q}\stackrel{\,g,\,G'}{\rightarrow} t\bar{t}, z<0)=  \nonumber
%  \\
%  &=&
%\Delta(\hat{s}) \; \frac{\overline{Y}}{|\overline{Y}|} \; \frac{f(p_q, p_{\bar{q}})}{I(p_q p_{\bar{q}})},
\Delta(\hat{s}) \; \frac{\overline{Y}}{|\overline{Y}|} \; \varkappa(p_q,p_{\bar{q}})
%\frac{f(p_q, p_{\bar{q}})}{ (p_q p_{\bar{q}}) }
%\nonumber
\label{DcG1qqbarttbar}
\end{eqnarray}
where
\begin{eqnarray}
%\\
%[3mm]
  \Delta(\hat{s}) &=& \frac{4\pi\beta^2 a^2}{9} \,
\frac{\alpha_s(\mu)\alpha_s(M_{chc})(\hat{s}-m_{G'}^2)+2\alpha_s^2(M_{chc})v^2 \hat{s}}
{(\hat{s}-m_{G'}^2)^2 + m_{G'}^2\Gamma_{G'}^2}
%\nonumber
\label{Dshat}
\end{eqnarray}
and
\begin{eqnarray}
%  f^{(\pm)}(z) &=& 1\pm\frac{4 m_t^2}{\hat{s}}+\mathrm{th}^2(z/2\overline{Y}),
%\nonumber
%\label{fpqpqbar}
\label{kappapqpqbar}
%\\
%f(p_q,p_{\bar{q}}) &=& p_{qz}\varepsilon_{\bar{q}}-p_{\bar{q}z}\varepsilon_q
\varkappa(p_q,p_{\bar{q}}) &=& (p_{qz}\varepsilon_{\bar{q}}-p_{\bar{q}z}\varepsilon_q)/(p_q p_{\bar{q}})
%\; \;\; \text{is antisymmetric under} \;\; q \leftrightarrow \bar{q}.
%\nonumber
\end{eqnarray}
is the antisymmetric under permutation $q \leftrightarrow \bar{q}$ function
of the momenta of the initial quark and antiquark.

As concerns the process $g g  \rightarrow t \bar{t}$
of $t \bar{t}$ production in gluon fusion the $G'$-boson does not contribute to this process
in tree approximation.

The total cross section of $t\bar{t}$-production in $pp$-collisions can be expressed in terms of
the parton cross sections $\sigma( i j \to t\bar{t})$ and the parton distribution functions
$f_i^p\!(x_1), \, f_j^p\!(x_2)$ in the usual way
\begin{eqnarray}
\label{csppttbar1}
\hspace{-6mm} &&  \sigma(pp\to t\bar{t}) =
\sum_{i,j} \iint f_i^p\!(x_1) f_j^p\!(x_2)\, \sigma( i j \to t\bar{t}) dx_1 dx_ ,
\hspace{6mm}  (i, j = q_k,\bar{q}_k, g).
%\nonumber
%  \\
%  i, j&=&q_k,\bar{q}_k, g,
% \nonumber
\end{eqnarray}

The parton cross sections $\sigma( i j \to t\bar{t})$ can be written as the sum of
the SM cross sections $\sigma^{\mathrm{SM}}( i j \to t\bar{t})$
and the contributions $\Delta \sigma^{G'}_{\mathrm{LO}}( i j \to t\bar{t})$ induced
in tree approximation by the $G'$-boson
%
%The parton cross sections as well as the total cross section of $t\bar{t}$-production
%can be written as the sum of the SM cross sections
%and the corresponding contributions induced in tree approximation by the $G'$-boson
%%
\begin{eqnarray}
  \label{eq:30}
  \sigma( i j \to t\bar{t})&=&
\sigma^{\mathrm{SM}}( i j \to t\bar{t})+\Delta \sigma^{G'}_{\mathrm{LO}}( i j \to t\bar{t}).
% \nonumber
%  \\
%  \Delta \sigma_{G'}^{\mathrm{LO}}(i,j \to t\bar{t}) &=&
%\Delta \sigma (q_k \bar{q}_k \stackrel{\,g,\, G'}{\rightarrow} t\bar{t}), \quad
%\mathrm{for}\quad i=q_k (\bar{q}_k),\, j=\bar{q}_k (q_k),
 \nonumber
\end{eqnarray}

The SM cross sections $\sigma^{\mathrm{SM}}( i j \to t\bar{t})$ can be written as the expansion
\begin{eqnarray}
%  \label{eq:31}
  \sigma^{\mathrm{SM}}( i j \to t\bar{t}) &=&
a_s^2 \, \Big[ \, \sigma_{\mathrm{SM}}^{(0)}( i j \to t \bar{t}) +
a_s \sigma_{\mathrm{SM}}^{(1)}( i j\to t\bar{t}) +
\nonumber
\\
&+&a_s^2\sigma_{\mathrm{SM}}^{(2)}( i j \to t \bar{t}) \, \Big] + O(a_s^5), \quad a_s=\alpha_s/\pi
 \nonumber
%\\
\end{eqnarray}
where $a_s^2 \, \sigma_{\mathrm{SM}}^{(0)}( i j \to t\bar{t})$
are the well known SM cross sections of tree approximation for
$ i=q_k (\bar{q}_k)g, \, j=\bar{q}_k (q_k)g $
and for the perturbation corrections %of the first two orders
we have used the expressions
$\sigma^{(1,2)}_{\mathrm{SM}}( i j \to t \bar{t}) = \hat{\sigma}_{i,j}^{(1,2)}(\hat{s},m_t,\mu_r,\mu_f)$
%$\hat{\sigma}_{i,j}^{(1,2)}(\hat{s},m_t,\mu_r,\mu_f)$
of ref.\cite{Aliev:2010zk}.

For the $G'$-boson contributions $\Delta \sigma^{G'}_{\mathrm{LO}}( i j \to t\bar{t})$
we use the expressions~\eqref{DsG1qqbarttbar}
%
%\label{DsG1qqbarttbar}
%
\begin{eqnarray}
%  \label{eq:30}
%  \sigma( i j \to t\bar{t})&=&
%\sigma_{\mathrm{SM}}( i j \to t\bar{t})+\Delta \sigma_{G'}^{\mathrm{LO}}( i j \to t\bar{t}),
% \nonumber
%  \\
  \Delta \sigma^{G'}_{\mathrm{LO}}( i j \to t\bar{t}) &=&
%\Delta \sigma (q_k \bar{q}_k \stackrel{\,g,\, G'}{\rightarrow} t\bar{t}),
\Delta \sigma_{LO}^{G'}(q_k \bar{q}_k  \rightarrow t \bar{t})
\quad \mathrm{\text{for}}\quad i=q_k (\bar{q}_k),\, j=\bar{q}_k (q_k).
 \nonumber
\end{eqnarray}

As a result we obtain the total cross section~\eqref{csppttbar1} as the sum
\begin{eqnarray}
  \label{csppttbar2}
  \sigma(pp \to t\bar{t}) &=&
\sigma^{\mathrm{SM}}(pp \to t\bar{t}) + \Delta \sigma^{G'}_{\mathrm{LO}}(pp \to t \bar{t})
% \nonumber
\end{eqnarray}
of the SM cross section $\sigma^{\mathrm{SM}}(pp \to t\bar{t})$
and the contribution $\Delta \sigma^{G'}_{\mathrm{LO}}(pp \to t \bar{t})$
induced in tree approximation by the $G'$-boson.

The charge difference of the parton cross section~\eqref{eq:DC} results in
the corresponding charge difference $\Delta_C(pp \to t \bar{t})$ of the $t\bar{t}$-production
in $pp$-collisions. One usually uses the charge asymmetry in the $t\bar{t}$-production % in $pp$-collisions
which we define as
\begin{eqnarray}
\hspace{-10mm} && A_C(p p \to t\bar{t}) =
\frac{ \sigma( p p \to t\bar{t}, z>0 )-\sigma( p p \to t\bar{t}, z<0 ) }
{\sigma( p p \to t\bar{t})} \equiv
%\nonumber
%\\
%&=&
\frac{ \Delta_C(pp \to t \bar{t}) }{\sigma( p p \to t\bar{t})}
\nonumber
%\label{eq:14}
\end{eqnarray}
(this definition coincides with the definition with use of the variable $\Delta |y|$ instead of $z$).

With account the $G'$-boson the charge asymmetry $A_C(p p \to t\bar{t})$ can be written as the sum
\begin{eqnarray}
  A_C(p p \to t\bar{t}) &=& A_C^{SM}(p p \to t\bar{t}) + A_C^{G'}(p p \to t\bar{t})
%\nonumber
\label{Acppttbar}
\end{eqnarray}
of the charge asymmetry $A_C^{SM}(p p \to t\bar{t})$ in the SM and of the contribution
induced by the $G'$-boson
\begin{eqnarray}
A_C^{G'}(p p \to t\bar{t}) &=&
\frac{ \Delta_C^{G'}(pp \stackrel{\,g,\,G'}{\rightarrow} t \bar{t}) }{\sigma( p p \to t\bar{t})}
%\frac{ \Delta_C^{LO}(pp \stackrel{\,g,\,G'}{\rightarrow} t \bar{t}) }{\sigma( p p \to t\bar{t})}
%\nonumber
\label{AcppttbarG1}
\end{eqnarray}
where
$\Delta_C^{G'}(pp \stackrel{\,g,\,G'}{\rightarrow} t \bar{t})$
is the contribution of the leading order to the charge difference $\Delta_C(pp \to t \bar{t})$
from the $G'$-boson.

The contribution of the $G'$-boson to the charge difference in the leading order
has been calculated with account of~\eqref{DcG1qqbarttbar},~\eqref{Dshat},~\eqref{kappapqpqbar} as
%
%\label{DcG1qqbarttbar}
%\label{Dshat}
%\label{fpqpqbar}
%\label{kappapqpqbar}
%
\begin{eqnarray}
%  \Delta_C^{LO}(pp\stackrel{\,g,\,G'}{\rightarrow}t\bar{t}) &=&
  \Delta_C^{G'}(pp\stackrel{\,g,\,G'}{\rightarrow}t\bar{t}) &=&
2\iint\limits_{D_1} F_{\Delta_C}^{pp}\!(x_1,x_2)\Delta(x_1 x_2 s) dx_1 dx_2
% \nonumber
%  \\
%[2mm]
%  s&=&(P_1+P_2)^2, \hspace{5mm}  \hat{s} = x_1 x_2 s.
%\nonumber
\label{DcppttbarG1}
\end{eqnarray}
where $\Delta(x_1 x_2 s)$ is defined by~\eqref{Dshat}
with $\hat{s} = x_1 x_2 s$, $ \; s = (P_1+P_2)^2$, \, $P_1, P_2$ are the momenta of the colliding protons.
The integration in~\eqref{DcppttbarG1} is performed over the region %$D_1$
\begin{eqnarray}
  D_1: \; \;  x_0^2/x_1\le x_2 \le x_1, \quad x_0\le x_1 \le 1,\quad  x_0^2 = 4m_t^2/s
 \nonumber
  \label{eq:26}
\end{eqnarray}
and the function $F_{\Delta_C}^{pp}\!(x_1,x_2)$ is defined by the parton distribution functions
of quarks and antiquarks as
\begin{eqnarray}
  F_{\Delta_C}^{pp}\!(x_1,x_2)=
\sum_k\left[f_{q_k}^p(x_1)f_{\bar{q}_k}^p(x_2)-
f_{\bar{q}_k}^p(x_1)f_{q_k}^p(x_2)\right].
%\nonumber
\label{FppDc}
\end{eqnarray}
The function~\eqref{FppDc} is antisymmetric under permutation $x_1 \leftrightarrow x_2$ and
is nonzero because of difference of the parton distribution functions
of the (valence) quarks and (see) antiquarks in proton (the minus sign appears due to the antisymmtric
function~\eqref{kappapqpqbar}, for $ m_q^2, m_p^2 \ll \, s $ this function takes
the values $\varkappa(p_q,p_{\bar{q}}) = \pm 1 $ in dependence on whether the quark flies along
the positive direction of $z$-axis (hence the antiquark flies in the opposite direction) or vice versa).

We have calculated and analysed the cross section~\eqref{csppttbar2} % $\sigma(pp \to t\bar{t})$
and the charge asymmetry~\eqref{Acppttbar}  %  $A_C(p p \to t\bar{t})$
in $t\bar{t}$-production in $pp$-collisions.
%
%We use the parton distribution functions MSTW2008   with $Q^2=\mu_r=\mu_f=m_t$ \cite{Martin:2009iq}
%and assume $\mu = m_t$.
%
For numerical calculations we use the parton cross sections with $\mu=\mu_R=m_t$ and
the parton distribution functions MSTW2008 \cite{Martin:2009iq}
with $\mu_F=m_t$.

For $\sigma^{\mathrm{SM}}(pp \to t\bar{t})$ as a result of calculation we have obtained the value
$  \sigma^{\mathrm{SM}}(pp \to t\bar{t}) = 162^{+7.8}_{-13.6} \, \mathrm{pb} $
%%
%\begin{eqnarray}
%  \sigma^{\mathrm{SM}}(pp \to t\bar{t}) &=& 162^{+7.8}_{-13.6} \, \mathrm{pb}
% \nonumber
%\end{eqnarray}
%%
in agreement with CMS experimental value~\eqref{csppttbarCMS} and with SM prediction~\eqref{csppttbarSM}.
%
%\cite{CMS-PAS-TOP-11-024} $\sigma_{t\bar{t}} = 165.8 \pm 13.3 \text{pb}$. %  \; [9].
%\begin{eqnarray}
%\hspace{-13mm} \sigma_{t\bar{t}} &=& 165.8 \pm 13.3 \text{pb}, \; [9].
%\nonumber
%\end{eqnarray}
%
The $G'$-boson contribution $\Delta \sigma^{G'}_{\mathrm{LO}}(pp \to t \bar{t})$
to the cross section~\eqref{csppttbar2} has been calculated by using
the formulas~\eqref{DsG1qqbarttbar},~\eqref{csppttbar1}  and occurs to be of a few pb in a reasonable
$m_{G'} - \theta_G$ region.

For the SM contribution $A_C^{SM}(p p \to t\bar{t})$ to the charge asymmetry
we have used the value~\eqref{AcppttbarSM}.
%of ref.~\cite{Kuhn:2011ri}
%%
%\begin{eqnarray}
%A_C^{SM}(p p \to t\bar{t}) &=& 0.0115~\mbox{\cite{Kuhn:2011ri}}.
%%%\pm 0.0006 % \, \cite{Kuhn2012}
%%\nonumber
%%\label{eq:14}
%\end{eqnarray}
%
The $G'$-boson contribution $A_C^{G'}(p p \to t\bar{t})$ to the charge asymmetry~\eqref{Acppttbar}
has been calculated by using the formulas~\eqref{AcppttbarG1},~\eqref{DcppttbarG1},~\eqref{FppDc}.

The $G'$-boson contributions $A_C^{G'}(p p \to t\bar{t})$ to the charge asymmetry
in $t \bar{t}$ production at the LHC
are shown in the Fig. 1 as the functions of the $G'$-boson mass~$m_{G'}$ for the mixing angles
$a) \, \theta_G = 45^\circ, \, b) \, \theta_G = 30^\circ, \,c) \, \theta_G = 15^\circ,$
\, ($\sqrt{s}= 7 \, \rm {TeV}$).
The dashed horizontal lines indicate the experimental values~\eqref{AcppttbarCMS} of the charge asymmetry
and the corresponding experimental errors of $1 \sigma$ and the solid horizontal line indicates
the SM prediction~\eqref{AcppttbarSM}.

\begin{figure}[htb]
\vspace*{-3mm}
\centerline{
\epsfysize=0.6
\textwidth
%\epsffile{AcG1LHC.pdf}}
%\epsffile{AcG1LHC.eps}}
\epsffile{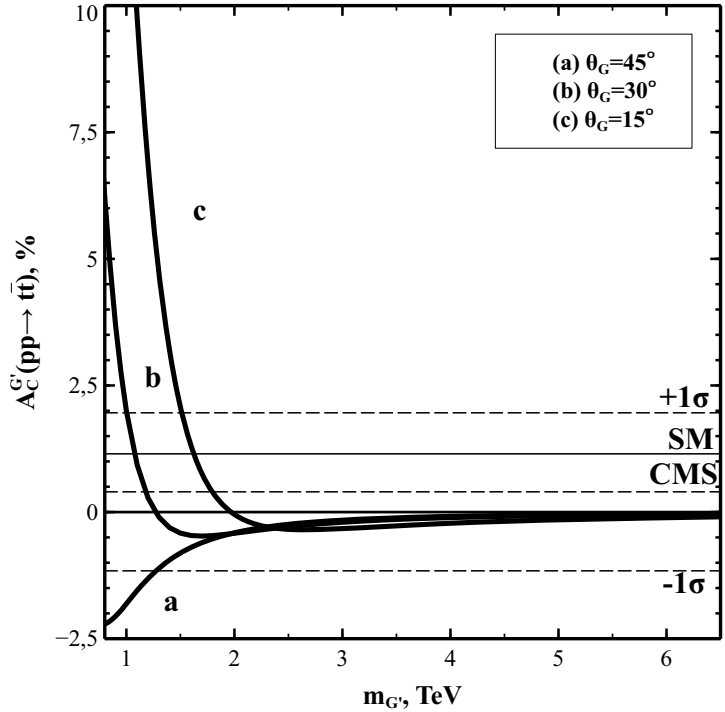}}
%\vspace*{-3mm}
%\label{fig:BrKS}
\caption{
The $G'$-boson contributions $A_C^{G'}(p p \to t\bar{t})$ to the charge asymmetry
in $t \bar{t}$ production at the LHC as the functions of the $G'$-boson mass~$m_{G'}$
in dependence on the mixing angle~$\theta_G$, \, ($\sqrt{s}= 7 \, \rm {TeV}$).
}
\end{figure}

%\label{AcppttbarCMS}
%\label{Acppttbar}
%\label{AcppttbarG1}
%
%\label{DcppttbarG1}
%\label{FppDc}

As seen from the Fig.1, for the $G'$-boson masses of order or larger 1 TeV and for the appropriate
mixing angles the $G'$-boson contributions $A_C^{G'}(p p \to t\bar{t})$ to the charge asymmetry
with account of the SM contribution~\eqref{AcppttbarSM} can be in agreement with CMS experimental
value~\eqref{AcppttbarCMS}.

\begin{figure}[htb]
\vspace*{-3mm}
\centerline{
\epsfysize=0.6
\textwidth
%\epsffile{limfcsAc_CMSCDF_SM_PMC2.pdf}}
%\epsffile{limfcsAc_CMSCDF_SM_PMC2.eps}}
\epsffile{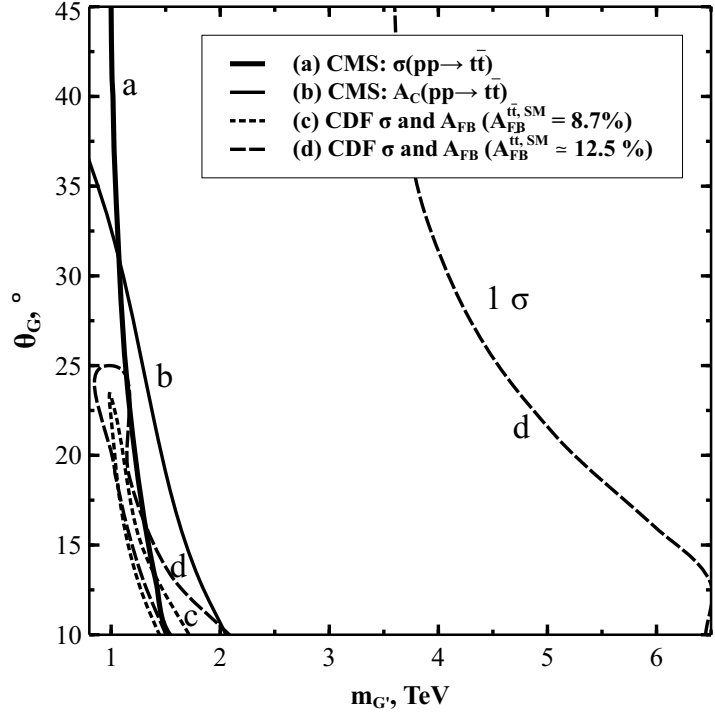}}
%\epsffile{csLHC.pdf}}
\vspace*{-3mm}
%\label{fig:BrKS}
\caption{
The $m_{G'} - \theta_G$ regions consistent within $1 \sigma$ with CMS data %\newline
$(a)$ on the cross section %\newline
$\sigma(pp \to t\bar{t})$ and %\newline
$(b)$ on the charge asymmetry~$A_C(p p \to t\bar{t})$ of $t \bar{t}$ production at the LHC, \,
($\sqrt{s}= 7 \, \rm {TeV}$)
and with CDF data on the cross section~$\sigma(p\bar{p} \to t\bar{t})$ and on the forward-backward
asymmetry $A_{\rm FB}(p\bar{p} \to t\bar{t})$ of $t \bar{t}$ production at the Tevatron for
c) $A_{FB}^{SM}(p\bar{p} \to t\bar{t}) = 8.7 \%$,    %  $\mathrm{\cite{Kuhn:2011ri}}$
d)$A_{FB}^{SM}(p\bar{p} \to t\bar{t}) =12.5 \%$.    %  $\mathrm{\cite{Brodsky:2012rj}}$.
}
\end{figure}

The regions in $m_{G'} - \theta_G$ plane which are consistent within $1 \sigma$ with CMS data
on the cross section $\sigma(pp \to t\bar{t})$~\eqref{csppttbarCMS} and
on the charge asymmetry~$A_C(p p \to t\bar{t})$~\eqref{AcppttbarCMS} are shown in the Fig.2,
\, ($\sqrt{s}= 7 \, \rm {TeV}$).
The curves $(a)$ and $(b)$ show the lower bounds of the $1 \sigma$ consistency regions resulted from
the cross section~\eqref{csppttbarCMS} and from the charge asymmetry~\eqref{AcppttbarCMS} respectively.
As seen from the Fig.2,
%the corresponding lower bounds on the $G'$-boson mass are of about $1.0 - 1.7 \, \, \text{TeV}$
the the $G'$-boson with masses
\begin{eqnarray}
m_{G'} > 1.0 - 1.7 \, \, \text{TeV}
\label{limmG1fAcCMS}
% \nonumber
\end{eqnarray}
for $\theta_G=45^\circ - 15^\circ$ respectively
is consistent within $1 \sigma$ with the CMS data~\eqref{csppttbarCMS},~\eqref{AcppttbarCMS}.

It should be noted however that $G'$-boson can have also the mass bounds from another sourses.
For example, $G'$-boson interacts also with the light quarks and could manifest itself as the peak in
the light quark dijet production. The unobservation of such peak at the LHC gives the corresponding
lower bound on $G'$-boson mass which can be more stringent than those resulting from $t \bar{t}$ production.
Nevertheless the bounds on $G'$-boson mass~\eqref{limmG1fAcCMS} obtained in this paper from
the $t \bar{t}$ production are interesting as independent ones.

The $G'$-boson gives rise also to the charge asymmetry in $p\bar{p} \to t\bar{t}$ process,
which results in the forward-backward asymmetry in $t \bar{t}$ production at the Tevatron
defined as
%
%$A_{\rm FB}(p\bar{p} \to t\bar{t})$ of $t \bar{t}$
%
\begin{eqnarray}
\hspace{-10mm} && A_{\rm FB}(p\bar{p} \to t\bar{t}) =
\frac{ \sigma( p \bar{p} \to t\bar{t}, \Delta y > 0 )-\sigma( p \bar{p} \to t\bar{t}, \Delta y < 0 ) }
{\sigma( p \bar{p} \to t\bar{t})} \equiv
%\nonumber
%\\
%&=&
\frac{ \Delta_{\rm FB}(p \bar{p} \to t \bar{t}) }{\sigma( p \bar{p} \to t\bar{t})}
%\nonumber
\label{AFBppbarttbarDef}
\end{eqnarray}
where $\Delta y = y_t - y_{\bar{t}}$ is related to the scattering angle $\hat{\theta}$ of $t$-quark
in the parton center of mass frame as $\mathrm{th}( \Delta y / 2) = \beta \cos \hat{\theta} $.

The region in $m_{G'} - \theta_G$ plane which is simultaneously
consistent  within $1 \sigma$ with CMS data on the cross section $\sigma(pp \to t\bar{t})$
and on the charge asymmetry~$A_C(p p \to t\bar{t})$
should be compared with the regions which are consistent within $1 \sigma$ with experimental data
on the cross section~$\sigma(p\bar{p} \to t\bar{t})$ and on the forward-backward
asymmetry $A_{\rm FB}(p\bar{p} \to t\bar{t})$ in $t \bar{t}$ production at the Tevatron.

At the parton level the $G'$-boson gives rise to the forward-backward difference
\begin{eqnarray}
\Delta_{\rm FB}(q\bar{q}\to t\bar{t}) = \sigma( q\bar{q}\to t\bar{t}, \Delta y>0 ) -
\sigma( q\bar{q}\to t\bar{t}, \Delta y<0 )
\nonumber
\label{Dfbqqbarttbar}
\end{eqnarray}
of $t \bar{t}$ production
in the form
\begin{eqnarray}
  \Delta_{\rm FB}^{G'}(q\bar{q}\stackrel{\,g,\,G'}{\rightarrow} t\bar{t}) &=&
%  \Delta_C^{LO}(q\bar{q}\stackrel{\,g,\,G'}{\rightarrow} t\bar{t}) &=&
%\sigma(q\bar{q}\stackrel{\,g,\,G'}{\rightarrow}t\bar{t}, z>0) -
%\sigma(q\bar{q}\stackrel{\,g,\,G'}{\rightarrow} t\bar{t}, z<0)=  \nonumber
%  \\
%  &=&
%\Delta(\hat{s}) \; \frac{\overline{Y}}{|\overline{Y}|} \; \frac{f(p_q, p_{\bar{q}})}{I(p_q p_{\bar{q}})},
\Delta(\hat{s}) \;  \varkappa(p_q,p_{\bar{q}})
%\frac{f(p_q, p_{\bar{q}})}{ (p_q p_{\bar{q}}) }
%\nonumber
\label{DfbG1qqbarttbar}
\end{eqnarray}
where $\Delta(\hat{s})$ and $\varkappa(p_q,p_{\bar{q}})$ are given
by equations~\eqref{Dshat},~\eqref{kappapqpqbar}.

With account of~\eqref{DfbG1qqbarttbar} the $G'$-boson contribution
$\Delta_{\rm FB}^{G'}(p \bar{p}\stackrel{\,g,\,G'}{\rightarrow}t\bar{t})$
to the forward-backward difference $\Delta_{\rm FB}(p \bar{p} \to t \bar{t}) $ in the leading order
can be expressed in terms of the parton distribution functions
of quarks and antiquarks in proton and antiproton as
%
%\label{DcG1qqbarttbar}
%\label{Dshat}
%\label{fpqpqbar}
%\label{kappapqpqbar}
%
\begin{eqnarray}
%  \Delta_C^{LO}(pp\stackrel{\,g,\,G'}{\rightarrow}t\bar{t}) &=&
\Delta_{\rm FB}^{G'}(p \bar{p} \stackrel{\,g,\,G'}{\rightarrow}  t\bar{t}) &=&
2\iint\limits_{D_1} F_{\Delta_{\rm FB}}^{p\bar{p}}\!(x_1,x_2)\Delta(x_1 x_2 s) dx_1 dx_2
% \nonumber
%  \\
%[2mm]
%  s&=&(P_1+P_2)^2, \hspace{5mm}  \hat{s} = x_1 x_2 s.
%\nonumber
\label{DfbppbarttbarG1}
\end{eqnarray}
where
\begin{eqnarray}
  F_{\Delta_{\rm FB}}^{p \bar{p} }\!(x_1,x_2)=
\sum_k\left[f_{q_k}^p(x_1)f_{\bar{q}_k}^{\bar{p}} (x_2)-
f_{\bar{q}_k}^p(x_1)f_{q_k}^{\bar{p}} (x_2)\right].
%\nonumber
\label{FpbarpDfb}
\end{eqnarray}
Because of the relations
$ f_{\bar{q}_k}^{\bar{p}}(x) = f_{q_k}^p(x), \, f_{q_k}^{\bar{p}}(x) = f_{\bar{q}_k}^p(x) $
the function~\eqref{FpbarpDfb} is symmetric under permutation $x_1 \leftrightarrow x_2$.
From~\eqref{AFBppbarttbarDef},~\eqref{DfbppbarttbarG1} we obtain
the $G'$-boson contribution
$A_{\rm FB}^{G'}(p\bar{p}\stackrel{\,g,\,G'}{\rightarrow}t\bar{t})$
to the forward-backward asymmetry in $t \bar{t}$ production at the Tevatron
in the form
\begin{eqnarray}
\hspace{-10mm} && A_{\rm FB}^{G'}(p\bar{p} \to t\bar{t}) =
\frac{ \Delta_{\rm FB}^{G'}(p \bar{p} \stackrel{\,g,\,G'}{\rightarrow}
t \bar{t}) }
{\sigma( p \bar{p} \to t\bar{t})}.
%\nonumber
\label{AFBppbarttbarG1}
\end{eqnarray}
%

%\label{AFBppbarttbarDef}
%\label{DfbppbarttbarG1}

For the further comparision we use below the CDF  data on $t\bar{t}$-production at the Tevatron
\begin{eqnarray}
\label{csppbarttbarCDF}
%
%\text{CDF}:
&& \sigma_{t\bar{t}} (p\bar{p} \to t\bar{t}) = 7.5 \pm 0.48  \,\text{pb} \, \mbox{\cite{CDF9913}},
%\cite{CDFCollab.2009}
%\nonumber
  \\
  && A_{FB}^{t\bar{t}} (p\bar{p} \to t\bar{t}) =0.164 \pm 0.045\,  \mbox{\cite{Aaltonen:2012it}}.
%\;\;\;\; ( 1.6 \sigma,  \, \text{above the}\, A_{FB}^{t\bar{t}} \, \text{of SM}).
%\cite{CDFCollab.2012}
\label{AFBppbarttbarCDF}
%\nonumber
%  \\
%\text{D0}:  && \sigma_{t\bar{t}} = 7.56_{-0.56}^{+0.63}(stat+syst) \text{pb} \,
% \fcite{Abazov2011403},
%%\cite{theD0Collaboration2011}
%\nonumber
% :  \\
% && A_{FB}^{t\bar{t}}=0.196 \pm 0.065\,   \fcite{Abazov:2011rq},
%\;\;\;\; ( 1.5 \sigma \, \text{above the}\, A_{FB}^{t\bar{t}} \, \text{of SM}).
%\cite{Abazov2011}
%\nonumber
\end{eqnarray}
and the current SM predictions
\begin{eqnarray}
\label{csppbarttbarSM}
%  && \sigma_{t\bar{t}}^{\text{NNLOapprox}}(m_t=173 \text{GeV}) = 7.08 \pm 0.36 \text{pb} \, [11]
  && \sigma_{t\bar{t}}^{\text{NNLOapprox}}(p\bar{p} \to t\bar{t}) = 7.08 \pm 0.36 \, \text{pb} \, \mbox{\cite{Kidonakis:2010dk}},
%\cite{Kidonakis2010}
%\nonumber
  \\
  && A_{FB}^{t\bar{t}} (p\bar{p} \to t\bar{t})  = 0.087(10) \,  \mbox{\cite{Kuhn:2011ri}},
%\;\;\;\; ( \text{} \,  1.6 \sigma  \, \text{below the CDF value~\eqref{AFBppbarttbarCDF}}),
%\;\;\;\; ( \text{на} \,  1.6 \sigma  \, \text{ниже значения CDF}),
% \; \text{(H. Kuhn, G. Rodrigo, 2012)}.
%\cite{Kuhn2012}
\label{AFBppbarttbarSM1}
%\nonumber
  \\
  && A_{FB}^{t\bar{t}} (p\bar{p} \to t\bar{t})  = 0.125 \,  \mbox{\cite{Brodsky:2012rj}}.
%\;\;\;\; ( \text{}   \, \text{consistent with the CDF value~\eqref{AFBppbarttbarCDF} within } \, 1 \sigma),
\label{AFBppbarttbarSM2}
%\nonumber
\end{eqnarray}
It should be noted that the SM value~\eqref{AFBppbarttbarSM1} of the forward-backward asymmetry
%$A_{\rm FB}(p\bar{p} \to t\bar{t})$
is  about $1.7 \sigma$ below the experimental value~\eqref{AFBppbarttbarCDF}
whereas the SM value~\eqref{AFBppbarttbarSM2} is consistent with~\eqref{AFBppbarttbarCDF}
within $1 \sigma$.

We have calculated the contributions of $G'$-boson to the cross section~$\sigma(p\bar{p} \to t\bar{t})$
and to the forward-backward asymmetry
$A_{\rm FB}(p\bar{p} \to t\bar{t})$ in $t \bar{t}$ production
at the Tevatron and have analysed them with account of the SM
contributions~\eqref{csppbarttbarSM},~\eqref{AFBppbarttbarSM1},~\eqref{AFBppbarttbarSM2}
in comparision with the experimental data~\eqref{csppbarttbarCDF},~\eqref{AFBppbarttbarCDF}.
We have found the regions in $m_{G'} - \theta_G$ plane which are
consistent within $1 \sigma$ with CDF data~\eqref{csppbarttbarCDF},~\eqref{AFBppbarttbarCDF}
with account of the SM predictions for the cross section~\eqref{csppbarttbarSM} and for
the forward-backward asymmetry~\eqref{AFBppbarttbarSM1}~or~\eqref{AFBppbarttbarSM2}.

%\label{csppbarttbarCDF}
%\label{AFBppbarttbarCDF}

The $G'$-boson contributions $A_{\rm FB}^{G'}(p \bar{p} \to t\bar{t})$
to the forward-backward asymmetry in $t \bar{t}$ production at the Tevatron
are shown in the Fig.~3 as the functions of the $G'$-boson mass~$m_{G'}$
for the mixing angles
$a) \, \theta_G = 45^\circ, \, b) \, \theta_G = 30^\circ, \,c) \, \theta_G = 15^\circ.$
The dashed horizontal lines indicate the experimental value~\eqref{AFBppbarttbarCDF}
and the corresponding experimental errors of $1 \sigma$ and the solid horizontal lines
indicate the SM predictions~\eqref{AFBppbarttbarSM1},~\eqref{AFBppbarttbarSM2}.

\begin{figure}[htb]
\vspace*{-3mm}
\centerline{
\epsfysize=0.6
\textwidth
%\epsffile{AfbG1Tevatron.pdf}}
%\epsffile{AfbG1Tevatron.eps}}
\epsffile{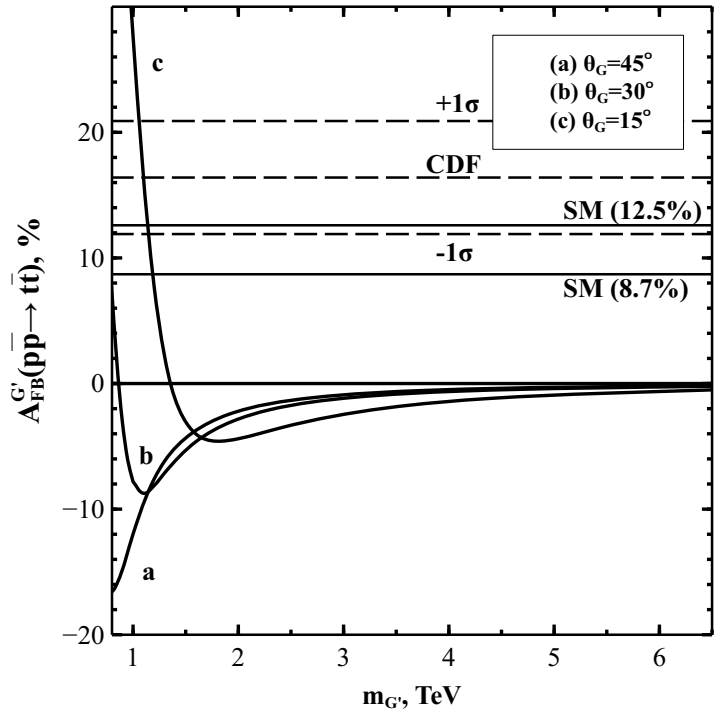}}
%\vspace*{-3mm}
%\label{fig:BrKS}
\caption{
The $G'$-boson contributions $A^{G'}_{FB}(p\bar{p} \to t\bar{t})$ to the forward-backward asymmetry
in $t \bar{t}$ production at the Tevatron as the functions of the $G'$-boson mass~$m_{G'}$
in dependence on the mixing angle~$\theta_G$.
}
\end{figure}

The $m_{G'} - \theta_G$ regions consistent within $1 \sigma$ with CDF data
on the cross section~$\sigma(p\bar{p} \to t\bar{t})$ and on the forward-backward
asymmetry $A_{\rm FB}(p\bar{p} \to t\bar{t})$ in $t \bar{t}$ production at the Tevatron
are shown in the Fig.~2 for the SM prediction~\eqref{AFBppbarttbarSM1} by dotted curve $c)$
and for the SM prediction~\eqref{AFBppbarttbarSM2} by dashed curve $d)$.

%for
%c) $A_{FB}^{SM}(p\bar{p} \to t\bar{t}) = 8.7 \%$,    %  $\mathrm{\cite{Kuhn:2011ri}}$
%d)$A_{FB}^{SM}(p\bar{p} \to t\bar{t}) =12.5 \%$.    %  $\mathrm{\cite{Brodsky:2012rj}}$.

As seen from the Fig.3, the $G'$-boson contribution $A_{\rm FB}^{G'}(p \bar{p} \to t\bar{t})$
to the forward-backward asymmetry in $t \bar{t}$ production at the Tevatron
for $ \theta_G = 45^\circ$(axigluon) is negative for all the $G'$-boson masses
but for $ \theta_G \sim 15^\circ - 20^\circ$ it can be positive
and for the $m_{G'} \sim 1.2 \, \rm {TeV}$ could improve the agreement of the SM
prediction~\eqref{AFBppbarttbarSM1} and the experimental value~\eqref{AFBppbarttbarCDF}
to $1 \sigma$ (the dotted region $c)$ in the Fig.2).
But, as seen from the Fig.2, the lower mass limit for $G'$-boson resulted from
CMS data~\eqref{AcppttbarCMS} on the charge asymmetry in $t \bar{t}$ production
at the LHC (the curve $b)$ )  exceeds  the $G'$-mass region $c)$ of $1 \sigma$ consistency
of the data~\eqref{AFBppbarttbarCDF} and the SM prediction~\eqref{AFBppbarttbarSM1}
of the forward-backward asymmetry in $t \bar{t}$ production at the Tevatron.
For $m_{G'} > 1.3 \, \rm {TeV}$ the $G'$-boson contributions $A_{\rm FB}^{G'}(p \bar{p} \to t\bar{t})$,
as also seen from the Fig.3,
are negative and for $m_{G'} \gtrsim 5 \, \rm {TeV}$ become sufficiently small.

For the  SM prediction~\eqref{AFBppbarttbarSM2} the region of $1 \sigma$ consistency
of the data~\eqref{AFBppbarttbarCDF} with the SM prediction~\eqref{AFBppbarttbarSM2}
near $m_{G'} \sim 1.2 \, \rm {TeV}$ in the Fig.2 is slightly larger
but still is below the curve $b)$.
But in this case there appears the allowed region of large $G'$-boson masses
with lower limis of order $m_{G'} = 3.5 - 6.0 \, \rm {TeV}$ (right dashed curve $d)$).
So, as seen from the Fig.2, in the case of the SM prediction~\eqref{AFBppbarttbarSM2}
the $G'$-boson with masses
\begin{eqnarray}
m_{G'} > 3.5 - 6.0 \, \, \text{TeV}
\label{limmG1fAFBCDF_SM_PC}
% \nonumber
\end{eqnarray}
for $\theta_G=45^\circ - 15^\circ$ respectively
is consistent within $1 \sigma$ with the CDF data~\eqref{csppbarttbarCDF},~\eqref{AFBppbarttbarCDF}
as well as with CMS data~\eqref{csppttbarCMS},~\eqref{AcppttbarCMS}.
It should be noted, that in the region of the masses and
mixing angles~\eqref{limmG1fAFBCDF_SM_PC}
%$m_{G'} > 3.5 - 6.0 \, \, \text{TeV}$
the $G'$-boson contribution to the charge asymmetry
in $t \bar{t}$ production at the LHC , as seen from the Fig.1, becomes small
and is of order of $-0.1 \%.$

%\label{csppbarttbarCDF}
%\label{AFBppbarttbarCDF}

In conclusion, we summarize the results found in this work.

The contributions of $G'$-boson predicted by the chiral color symmetry of quarks
to the charge asymmetry $A_C(p p \to t\bar{t})$ in $t \bar{t}$ production at the LHC
and to the forward-backward asymmetry $A_{\rm FB}(p\bar{p} \to t\bar{t})$
in $t\bar{t}$ production at the Tevatron
are calculated and analysed in dependence on two free parameters of the model,
the $G'$ mass $m_{G'}$ and mixing angle $\theta_G$.

The $m_{G'} - \theta_G$ regions of $1 \sigma$ consistency with the CMS data on
the cross section $\sigma(pp \to t\bar{t})$ and on the charge asymmetry $A_C(p p \to t\bar{t})$
are found and compared with those resulted from the CDF data
on the cross section~$\sigma(p\bar{p} \to t\bar{t})$
and on the forward-backward asymmetry~$A_{\rm FB}(p\bar{p} \to t\bar{t})$
of $t \bar{t}$ production at the Tevatron
with account of the current SM predictions for~$A_{\rm FB}(p\bar{p} \to t\bar{t})$.

It is shown that in the case of the SM prediction~\eqref{AFBppbarttbarSM1}
the $m_{G'} - \theta_G$ region of $1 \sigma$ consistency with the CMS data on
 the charge asymmetry $A_C(p p \to t\bar{t})$ exceeds
that resulted from the CDF data
on the forward-backward asymmetry $A_{\rm FB}(p\bar{p} \to t\bar{t})$
in $t\bar{t}$ production at the Tevatron.

In the case of the SM prediction~\eqref{AFBppbarttbarSM2}
the $G'$-boson with masses
%~\eqref{limmG1fAFBCDF_SM_PC}
$m_{G'} > 3.5 - 6.0 \, \, \text{TeV}$
%%
%\begin{eqnarray}
%m_{G'} > 3.5 - 6.0 \, \, \text{TeV}
%\label{limmG1fAFBCDF_SM_PC}
%% \nonumber
%\end{eqnarray}
%%
for $\theta_G=45^\circ - 15^\circ$ %respectively
is shown  to be consistent within $1 \sigma$ %both
with the CDF data
%~\eqref{csppbarttbarCDF},~\eqref{AFBppbarttbarCDF}
on $\sigma(p\bar{p} \to t\bar{t})$, $A_{\rm FB}(p\bar{p} \to t\bar{t})$
and with the CMS data
on $\sigma(pp \to t\bar{t})$, $A_C(p p \to t\bar{t})$
%~\eqref{csppttbarCMS},~\eqref{AcppttbarCMS}
simultaneously.

The work is supported by the Ministry of Education and Science of Russia
under state contract No.P795
of the Federal Programme "Scientific and Pedagogical Personnel of Innovation Russia"
for 2009-2013 years.

\bibliographystyle{science}
\bibliography{my_base2}

\end{document}